\documentclass[aps,prl,linenumbers,reprint,superscriptaddress]{revtex4-1}
\usepackage{graphicx}
\usepackage{multirow}

\begin{document}

\title{Double diffractive cross-section measurement in the forward region at LHC}

\author{G.~Antchev}
\affiliation{INRNE-BAS, Institute for Nuclear Research and Nuclear Energy, Bulgarian Academy of Sciences, Sofia, Bulgaria.}
\author{P.~Aspell}
\affiliation{CERN, Geneva, Switzerland.}
\author{I.~Atanassov}
\affiliation{CERN, Geneva, Switzerland.}
\affiliation{INRNE-BAS, Institute for Nuclear Research and Nuclear Energy, Bulgarian Academy of Sciences, Sofia, Bulgaria.}
\author{V.~Avati}
\author{J.~Baechler}
\affiliation{CERN, Geneva, Switzerland.}
\author{V.~Berardi}
\affiliation{INFN Sezione di Bari, Bari, Italy.}
\affiliation{Dipartimento Interateneo di Fisica di Bari, Bari, Italy.}
\author{M.~Berretti}
\author{E.~Bossini}
\affiliation{INFN Sezione di Pisa, Pisa, Italy.}
\affiliation{Universit\`{a} degli Studi di Siena and Gruppo Collegato INFN di Siena, Siena, Italy.}
\author{U.~Bottigli}
\affiliation{Universit\`{a} degli Studi di Siena and Gruppo Collegato INFN di Siena, Siena, Italy.}
\author{M.~Bozzo}
\affiliation{INFN Sezione di Genova, Genova, Italy.}
\affiliation{Universit\`{a} degli Studi di Genova, Genova, Italy.}
\author{E.~Br\"{u}cken}
\affiliation{Helsinki Institute of Physics, Helsinki, Finland.}
\affiliation{Department of Physics, University of Helsinki, Helsinki, Finland.}
\author{A.~Buzzo}
\affiliation{INFN Sezione di Genova, Genova, Italy.}
\author{F.~S.~Cafagna}
\affiliation{INFN Sezione di Bari, Bari, Italy.}
\author{M.~G.~Catanesi}
\affiliation{INFN Sezione di Bari, Bari, Italy.}
\author{M.~Csan\'{a}d}
\altaffiliation{Department of Atomic Physics, E\"otv\"os University, Budapest, Hungary.}
\affiliation{MTA Wigner Research Center, RMKI, Budapest, Hungary.}
\author{T.~Cs\"{o}rg\H{o}}
\affiliation{MTA Wigner Research Center, RMKI, Budapest, Hungary.}
\author{M.~Deile}
\affiliation{CERN, Geneva, Switzerland.}
\author{K.~Eggert}
\affiliation{Case Western Reserve University, Dept. of Physics, Cleveland, OH 44106, USA.}
\author{V.~Eremin}
\affiliation{Ioffe Physical-Technical Institute of Russian Academy of Sciences, St Petersburg, Russian Federation}
\author{F.~Ferro}
\affiliation{INFN Sezione di Genova, Genova, Italy.}
\author{A. Fiergolski}
\affiliation{Warsaw University of Technology, Warsaw, Poland.}
\affiliation{INFN Sezione di Bari, Bari, Italy.}
\author{F.~Garcia}
\affiliation{Helsinki Institute of Physics, Helsinki, Finland.}
\author{S.~Giani}
\affiliation{CERN, Geneva, Switzerland.}
\author{V.~Greco}
\affiliation{Universit\`{a} degli Studi di Siena and Gruppo Collegato INFN di Siena, Siena, Italy.}
\author{L.~Grzanka}
\altaffiliation{Institute of Nuclear Physics, Polish Academy of Science, Krakow, Poland.}
\affiliation{CERN, Geneva, Switzerland.}
\author{J.~Heino}
\affiliation{Helsinki Institute of Physics, Helsinki, Finland.}
\author{T.~Hilden}
\affiliation{Helsinki Institute of Physics, Helsinki, Finland.}
\affiliation{Department of Physics, University of Helsinki, Helsinki, Finland.}
\author{A.~Karev}
\affiliation{CERN, Geneva, Switzerland.}
\author{J.~Ka\v{s}par}
\author{J.~Kopal}
\affiliation{Institute of Physics of the Academy of Sciences of the Czech Republic, Praha, Czech Republic.}
\affiliation{CERN, Geneva, Switzerland.}
\author{V.~Kundr\'{a}t}
\affiliation{Institute of Physics of the Academy of Sciences of the Czech Republic, Praha, Czech Republic.}
\author{K.~Kurvinen}
\affiliation{Helsinki Institute of Physics, Helsinki, Finland.}
\author{S.~Lami}
\affiliation{INFN Sezione di Pisa, Pisa, Italy.}
\author{G.~Latino}
\affiliation{Universit\`{a} degli Studi di Siena and Gruppo Collegato INFN di Siena, Siena, Italy.}
\author{R.~Lauhakangas}
\affiliation{Helsinki Institute of Physics, Helsinki, Finland.}
\author{T.~Leszko}
\affiliation{Warsaw University of Technology, Warsaw, Poland.}
\author{E.~Lippmaa}
\affiliation{National Institute of Chemical Physics and Biophysics NICPB, Tallinn, Estonia.}
\author{J.~Lippmaa}
\affiliation{National Institute of Chemical Physics and Biophysics NICPB, Tallinn, Estonia.}
\author{M.~Lokaj\'{\i}\v{c}ek}
\affiliation{Institute of Physics of the Academy of Sciences of the Czech Republic, Praha, Czech Republic.}
\author{L.~Losurdo}
\affiliation{Universit\`{a} degli Studi di Siena and Gruppo Collegato INFN di Siena, Siena, Italy.}
\author{M.~Lo~Vetere}
\affiliation{INFN Sezione di Genova, Genova, Italy.}
\affiliation{Universit\`{a} degli Studi di Genova, Genova, Italy.}
\author{F.~Lucas~Rodr\'{i}guez}
\affiliation{CERN, Geneva, Switzerland.}
\author{M.~Macr\'{\i}}
\affiliation{INFN Sezione di Genova, Genova, Italy.}
\author{T.~M\"aki}
\affiliation{Helsinki Institute of Physics, Helsinki, Finland.}
\author{A.~Mercadante}
\affiliation{INFN Sezione di Bari, Bari, Italy.}
\author{N.~Minafra} 
\affiliation{INFN Sezione di Bari, Bari, Italy.}
\affiliation{Dipartimento Interateneo di Fisica di Bari, Bari, Italy.}
\author{S.~Minutoli}
\affiliation{CERN, Geneva, Switzerland.}
\affiliation{INFN Sezione di Genova, Genova, Italy.}
\author{F.~Nemes}
\altaffiliation{Department of Atomic Physics, E\"otv\"os University, Budapest, Hungary.}
\affiliation{MTA Wigner Research Center, RMKI, Budapest, Hungary.}
\author{H.~Niewiadomski}
\affiliation{CERN, Geneva, Switzerland.}
\author{E.~Oliveri}
\affiliation{Universit\`{a} degli Studi di Siena and Gruppo Collegato INFN di Siena, Siena, Italy.}
\author{F.~Oljemark}
\author{R.~Orava}
\affiliation{Helsinki Institute of Physics, Helsinki, Finland.}
\affiliation{Department of Physics, University of Helsinki, Helsinki, Finland.}
\author{M.~Oriunno}
\altaffiliation{SLAC National Accelerator Laboratory, Stanford CA, USA.}
\affiliation{CERN, Geneva, Switzerland.}
\author{K.~\"{O}sterberg}
\affiliation{Helsinki Institute of Physics, Helsinki, Finland.}
\affiliation{Department of Physics, University of Helsinki, Helsinki, Finland.}
\author{P.~Palazzi}
\affiliation{Universit\`{a} degli Studi di Siena and Gruppo Collegato INFN di Siena, Siena, Italy.}
\author{J.~Proch\'{a}zka}
\affiliation{Institute of Physics of the Academy of Sciences of the Czech Republic, Praha, Czech Republic.}
\author{M.~Quinto}
\affiliation{INFN Sezione di Bari, Bari, Italy.}
\affiliation{Dipartimento Interateneo di Fisica di Bari, Bari, Italy.}
\author{E.~Radermacher}
\affiliation{Universit\`{a} degli Studi di Siena and Gruppo Collegato INFN di Siena, Siena, Italy.}
\author{E.~Radicioni}
\affiliation{INFN Sezione di Bari, Bari, Italy.}
\author{F.~Ravotti}
\affiliation{CERN, Geneva, Switzerland.}
\author{E.~Robutti}
\affiliation{INFN Sezione di Genova, Genova, Italy.}
\author{L.~Ropelewski}
\author{G.~Ruggiero}
\affiliation{CERN, Geneva, Switzerland.}
\author{H.~Saarikko}
\affiliation{Helsinki Institute of Physics, Helsinki, Finland.}
\affiliation{Department of Physics, University of Helsinki, Helsinki, Finland.}
\author{A.~Scribano}
\affiliation{Universit\`{a} degli Studi di Siena and Gruppo Collegato INFN di Siena, Siena, Italy.}
\author{J.~Smajek}
\author{W.~Snoeys}
\affiliation{CERN, Geneva, Switzerland.}
\author{J.~Sziklai}
\affiliation{MTA Wigner Research Center, RMKI, Budapest, Hungary.}
\author{C.~Taylor}
\affiliation{Case Western Reserve University, Dept. of Physics, Cleveland, OH 44106, USA.}
\author{N.~Turini}
\affiliation{Universit\`{a} degli Studi di Siena and Gruppo Collegato INFN di Siena, Siena, Italy.}
\author{V.~Vacek}
\author{M.~V\'itek}
\affiliation{Czech Technical University, Praha, Czech Republic.}
\author{J.~Welti} 
\affiliation{Helsinki Institute of Physics, Helsinki, Finland.}
\affiliation{Department of Physics, University of Helsinki, Helsinki, Finland.}
\author{J.~Whitmore}  
\affiliation{Penn State University, Dept.~of Physics, University Park, PA 16802, USA.}
\author{P.~Wyszkowski}
\altaffiliation{AGH University of Science and Technology, Krakow, Poland.}
\affiliation{CERN, Geneva, Switzerland.}

\collaboration{The TOTEM Collaboration}

\vspace{0.5cm}

\date{\today}

\begin{abstract}
The first double diffractive cross-section measurement in the very forward region has been carried out by the TOTEM experiment at the LHC with center-of-mass energy of $\sqrt{s}=7$~TeV. By utilizing the very forward TOTEM tracking detectors T1 and T2, which extend up to $|\eta|$=6.5, a clean sample of double diffractive pp events was extracted. From these events, we measured the cross-section $\sigma_{\rm DD}=({116 \pm 25})\,~{\rm \mu b}$ for events where both diffractive systems have 4.7$<$$|\eta|_{min}$$<$6.5. 
\end{abstract}


\maketitle

Diffractive scattering represents a unique tool for investigating the dynamics of strong interactions and proton structure. These events are dominated by soft processes which cannot be calculated with perturbative QCD. Various model calculations predict diffractive cross-sections that are markedly different at the LHC energies~\cite{theory1,theory2,theory3}.

Double diffraction (DD) is the process in which two colliding hadrons dissociate into clusters of particles, and the interaction is mediated by an object with the quantum numbers of the vacuum. Experimentally, DD events are typically associated with a rapidity gap that is large compared to random multiplicity fluctuations. Rapidity gaps are exponentially suppressed in non-diffractive (ND) events~\cite{rapgap}, however when a detector is not able to detect particles with the transverse momentum ($p_{T}$) of a few hundred~MeV, the identification of double diffractive events by means of rapidity gaps becomes very challenging. The excellent $p_{T}$ acceptance of the TOTEM detectors makes the experiment favorable for the measurement. Previous measurements of DD cross-section are described in~\cite{cdf,alice}.

The TOTEM experiment \cite{totem} is a dedicated experiment to study diffraction, total cross-section and elastic scattering at the LHC. It has three subdetectors placed symmetrically on both sides of the interaction point: Roman Pot detectors to identify leading protons and T1 and T2 telescopes to detect charged particles in the forward region. The most important detectors for this measurement are the T2 and T1 telescopes. T2 consists of Gas Electron Multipliers that detect charged particles with $p_{T}>$40~MeV/c at pseudo-rapidities of 5.3$<$$|\eta|$$<$6.5 \cite{pseudorapidity}. The T1 telescope consists of Cathode Strip Chambers that measure charged particles with $p_{T}>$100~MeV/c at 3.1$<$$|\eta|$$<$4.7.

In this novel measurement, the double diffractive cross-section was determined in the forward region. The method is as model-independent as possible. The DD events were selected by vetoing T1 tracks and requiring tracks in T2, hence selecting events that have two diffractive systems with 4.7$<$$|\eta|_{min}$$<$6.5, where $\eta_{min}$ is the minimum pseudorapidy of all primary particles produced in the diffractive system. Although these events are only about 3$\%$ of the total $\sigma_{DD}$, they provide a pure selection of DD events and the measurement is an important step towards determining if there is a rich resonance structure in the low mass region~\cite{resonances}. To probe further, the $\eta_{min}$ range was divided into two sub-regions on each side, providing four subcategories for the measurement.

The analysis is structured in three steps. In the first step, the raw rate of double diffractive events is estimated: the selected sample is corrected for trigger efficiency, pile-up and T1 multiplicity, and the amount of background is determined. In the second step, the visible cross-section is calculated by correcting the raw rate for acceptance and efficiency to detect particles. In the last step, the visible cross-section is corrected so that both diffractive systems have 4.7$<$$|\eta|_{min}$$<$6.5. 

This measurement uses data collected in October 2011 at $\sqrt{s}$=7~TeV during a low pile-up run with a special $\beta^*$=90 m optics. The data were collected with the T2 minimum bias trigger. The trigger condition was that 3 out of 10 superpads in the same $r-\phi$ sector fired. A superpad consists of 3 radial and 5 azimuthal neighbouring pads, and it is sufficient that one out of 15 pads registered a signal for a superpad to be fired.


After the offline reconstruction~\cite{offline}, the DD events were selected by requiring tracks in both T2 arms and no tracks in either of the T1 arms (2T2+0T1). T2 tracks with a $\chi^2$-fit probability smaller than 2$\%$ and tracks falling in the overlap region of two T2 quarters, i.e. tracks with 80$^\circ$$<$$\phi$$<$100$^\circ$ or 260$^\circ$$<$$\phi$$<$280$^\circ$, were removed. The tracks in the overlap region were removed because simulation does not model well their response. In the paper, this full selection for visible cross-section is named ${\rm I_{track}}$. The four subcategories for the visible cross-section measurement were defined by the T2 track with minimum $|\eta|$ on each side, $|\eta^{+}_{track}|_{min}$ and $|\eta^{-}_{track}|_{min}$. The subcategory ${\rm D11_{track}}$ includes the events with  5.3$<$$|\eta^{\pm}_{track}|_{min}$$<$5.9, ${\rm D22_{track}}$ the events with 5.9$<$$|\eta^{\pm}_{track}|_{min}$$<$6.5, ${\rm D12_{track}}$ the events with  5.3$<$$|\eta^{+}_{track}|_{min}$$<$5.9 and 5.9$<$$|\eta^{-}_{track}|_{min}$$<$6.5, and ${\rm D21_{track}}$ the events with  5.9$<$$|\eta^{+}_{track}|_{min}$$<$6.5 and 5.3$<$$|\eta^{-}_{track}|_{min}$$<$5.9.

Two additional samples were extracted for background estimation. A control sample for single diffractive (SD) events has at least one track in either of the T2 arms and no tracks in the opposite side T2 arm nor in T1 (1T2+0T1). A control sample for ND events has tracks in all arms of T2 and T1 detectors (2T2+2T1). Four additional exclusive data samples were defined for testing the background model validity: tracks in both arms of T2 and exactly in one arm of T1 (2T2+1T1), tracks in either of T2 arms and in both T1 arms (1T2+2T1), tracks in T2 and T1 in one side of the interaction point (1T2+1T1 same side) and tracks in T2 and T1 in the opposite side of the interaction point (1T2+1T1 opposite side). Each sample corresponds to one signature type $j$.

The number of selected data events was corrected for trigger efficiency and pile-up. The trigger efficiency correction $c_{t}$ was calculated from zero-bias triggered sample in the bins of number of tracks. It is described in detail in \cite{inelastic}. The pile-up correction was calculated using the formula:
\begin{linenomath}
\begin{equation}
c_{pu}^{j}=\frac{1}{1-\frac{2 p_{pu}}{1+p_{pu}}+\frac{2 p_{pu}}{1+p_{pu}} \cdot p^{j}}
\end{equation}
\end{linenomath}
where $j$ is the signature type, $p_{pu}$=(1.5$\pm$0.4)$\%$ is the pile-up correction factor for inelastic events \cite{inelastic}, and $p^{j}$ is the correction for signature type changes due to pile-up. The correction $p^{j}$ was determined by creating a MC study of pile-up. A pool of signature types was created by weighting each type with their probability in the data. Then a pair was randomly selected, and their signatures were combined. After repeating the selection and combination, the correction was calculated as $p^{j}$=$N_{combined}^{j}/N_{original}^{j}$. $N_{combined}^{j}$ is the number of selected combinations that have the combined signature of $j$. The uncertainty in $p^{j}$ was determined by taking the event type weights from Pythia~8~\cite{pythia} and recalculating $p^{j}$. The corrected number of data events were calculated with the formula $N^{j}=c_{t} c_{pu}^{j} N_{raw}^{j}$.

The simulated T1 track multiplicity distribution predicts a lower number of zero-track events than what was observed in the data. The number of T1 tracks in the simulation was corrected to match with the data by randomly selecting 10$\%$ (2$\%$) of one-(two-)track events and changing them to zero-track events.

Three kinds of background were considered for the analysis: ND, SD and central diffraction (CD). ND and SD background estimation methods were developed to minimize the model dependence, and the values of estimates were calculated iteratively. Since the CD background is significantly smaller than the ND and SD ones, its estimate ($N_{CD}$) was taken from simulation, using the acceptance and $\sigma_{CD}$=1.3~mb from Phojet~\cite{phojet}. 

The number of ND events in the ND dominated control sample, 2T2+2T1, has been determined as:
\begin{linenomath}
\begin{equation}
N_{ND}^{2T2+2T1}=N_{data}^{2T2+2T1}-N_{DD}^{2T2+2T1}-N_{SD}^{2T2+2T1}-N_{CD}^{2T2+2T1},
\end{equation} 
\end{linenomath}
where $N_{DD}^{2T2+2T1}$ and $N_{SD}^{2T2+2T1}$ were taken from MC for the first iteration. Pythia was used as the default generator throughout the analysis. The ratio, $R_{ND}^{j}$, of ND events expected in the sample $j$ and in the control sample, was calculated from MC as
\begin{linenomath}
\begin{equation}
R_{ND}^{j}=\frac{N_{ND,MC}^{j}}{N_{ND,MC}^{2T2+2T1}}.
\end{equation} 
\end{linenomath}
The number of ND events within the signal sample was estimated as
\begin{linenomath}
\begin{equation}
N_{ND}^{j}=R_{ND}^{j}\cdot C^{j} \cdot N_{ND}^{2T2+2T1},
\end{equation}
\end{linenomath}
where $C^{j}$ is the normalization factor deduced from the relative mismatch between the data and the total Pythia prediction in the signal sample:
\begin{linenomath}
\begin{equation}
C^{j}=\frac{N_{data}^{j}}{N_{MC}^{j}}\cdot \frac{N_{MC}^{2T2+2T1}}{N_{data}^{2T2+2T1}}.
\end{equation}
\end{linenomath}

The SD background estimation starts from the calculation of the number of SD events in the SD dominated control sample, 1T2+0T1, by subtracting the number of other kind of events from the number of data events:
\begin{linenomath}
\begin{equation}
N_{SD}^{1T2+0T1}=N_{data}^{1T2+0T1}-N_{DD}^{1T2+0T1}-N_{ND}^{1T2+0T1}-N_{CD}^{1T2+0T1},
\end{equation}
\end{linenomath}
where $N_{ND}^{1T2+0T1}$ was calculated with the ND estimation method and $N_{DD}^{1T2+0T1}$ was taken from Pythia for the first iteration. To scale the number of SD events to the signal region, the ratio $R_{SD}^{j}$ was calculated from data. The SD dominated data events that were used in the calculation of the ratio have exactly one leading proton seen by the RPs, in addition to the sample selections based on T2 and T1 tracks. By using the ratio
\begin{linenomath}
\begin{equation}
R_{SD}^{j}=\frac{N_{data}^{j+1 proton}}{N_{data}^{1T2+0T1+1 proton}},
\end{equation}
\end{linenomath}
the expected number of background SD events was calculated as 
\begin{linenomath}
\begin{equation}
N_{SD}^{j}=R_{SD}^{j}\cdot N_{SD}^{1T2+0T1}.
\end{equation}
\end{linenomath}

The first estimate of $\sigma_{DD}$ was calculated with the ND, SD and CD background estimates described above. The background estimations were repeated with redefined values of $N_{DD}^{2T2+2T1}$, $N_{SD}^{2T2+2T1}$, $N_{DD}^{1T2+0T1}$, $N_{ND}^{1T2+0T1}$: the numbers of DD events were scaled with the ratio of $\sigma_{DD}^{measured}/\sigma_{DD}^{MC}$, and the numbers of SD and ND events were calculated using their estimation methods. Next, the three steps were repeated until $N_{ND}^{2T2+0T1}$ and $N_{SD}^{2T2+0T1}$ converged. The final numbers of estimates in the ${\rm I_{track}}$ control samples are shown in Table~\ref{control}, and the estimated numbers of background events in the signal sample are shown in Table~\ref{bckg}.

\begin{table}
\caption{Estimated numbers of ND, SD, CD and DD events in the ND and SD background control samples. The numbers correspond to the full selection ${\rm I_{track}}$.
\label{control}}
\begin{ruledtabular}
\begin{tabular}{lcc}
& ND control sample & SD control sample \\
& 2T2+2T1 & 1T2+0T1 \\
\hline
ND & 1,178,737$\pm$19,368 & 659$\pm$65 \\
SD & 74,860$\pm$6,954 & 60,597$\pm$12,392 \\
CD & 2,413$\pm$1,207 & 2,685$\pm$1,343 \\
DD & 54,563$\pm$19,368 & 15,858$\pm$1,123 \\
\hline
Total & 1,310,573$\pm$20,614 & 79,798$\pm$12,465 \\
\hline
Data & 1,310,573 & 79,798 \\
\end{tabular}
\end{ruledtabular}
\end{table}

\begin{figure}
\includegraphics[width=90mm]{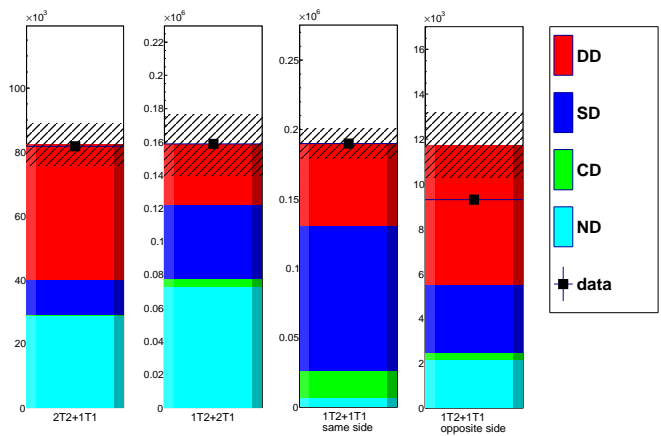}%
\caption{Validation of background estimates for the full selection ${\rm I_{track}}$. Each plot shows the corrected number of events in data (black squares) and the combined estimate with background uncertainties. The combined estimate is the sum of ND estimate (cyan), CD estimate (green), SD estimate (blue) and DD estimate (red). The shaded area represents the total uncertainty of the background estimate.\label{figure}}
\end{figure}

The reliability of the background estimates was examined in the validation samples. In these samples, the total estimated number of events is consistent with the number of data events within the uncertainty of the estimate, see Figure~\ref{figure}. The uncertainty in the SD estimate was determined with an alternative control sample: 1T2+1T1 same side. To determine the uncertainty in the ND estimate, the ratio $R_{ND}^{j}$ was calculated from Phojet and $N_{ND}^{j}$ estimated with it. A conservative uncertainty of 50$\%$ was assigned for the CD estimate.

The visible DD cross-section was calculated using the formula
\begin{linenomath}
\begin{equation}
\sigma_{DD}=\frac{E\cdot (N_{data}^{2T2+0T1}-N_{bckg}^{2T2+0T1})}{\mathcal{L}}
\end{equation}
\end{linenomath}
where $E$ is the experimental correction and the integrated luminosity $\mathcal{L}$=(40.1$\pm$1.6)~$\mu b^{-1}$. The experimental correction includes the acceptance, the tracking and reconstruction efficiencies of T2 and T1 detectors, the fraction of events with only neutral particles within detector acceptance, and bin migration. The correction was estimated using Pythia, and the largest difference with respect to QGSJET-II-03 \cite{qgsjet} and Phojet was taken as the uncertainty. An additional correction was introduced for the selections with 5.9$<$$|\eta_{track}|_{min}$$<$6.5 to scale the ratio $N_{5.9<|\eta_{track}|_{min}<6.5}/N_{total}$ to be consistent with data. 2T2+2T1 and 1T2+1T1 same side selections were used to achieve the scale factor. The value of the additional correction is 1.22$\pm$0.03 (1.24$\pm$0.03) for the positive (negative) side.

The visible cross-section was then corrected to the true $\eta_{min}$ cross-section. Pythia and Phojet predict a significantly different share of visible events that have their true $\eta_{min}$ within the uninstrumented region of 4.7$<$$|\eta|$$<$5.3. Therefore, the visible $\eta$ range was extended to $|\eta|$=4.7 to minimize the model dependence. This final correction was determined from generator level Pythia by calculating the ratio of $N_{4.7<|\eta^{\pm}|_{min}<6.5}/N_{visible}$. The uncertainty was estimated by comparing the nominal correction to the one derived from Phojet. In this paper, the true $\eta_{min}$ corrected cross-section (4.7$<$$|\eta^{\pm}|_{min}$$<$6.5) is called I, and the subcategories as D11 (4.7$<$$|\eta^{\pm}|_{min}$$<$5.9), D22 (5.9$<$$|\eta^{\pm}|_{min}$$<$6.5), D12 (4.7$<$$|\eta^{+}|_{min}$$<$5.9 and 5.9$<$$|\eta^{-}|_{min}$$<$6.5), and D21 (5.9$<$$|\eta^{+}|_{min}$$<$6.5 and 4.7$<$$|\eta^{-}|_{min}$$<$5.9).

The sources and values of systematic uncertainties are summarized in Table~\ref{systematics}. For each source of systematic uncertainty, the value was calculated by varying the source within its uncertainty and recalculating the measured cross-section. The difference between the nominal and recalculated cross-section was taken as the systematic uncertainty. 

In summary, we have measured the DD cross-section in an $\eta$ range where it has never been determined before. The TOTEM measurement is  $\sigma_{DD}$=(116$\pm$25)~$\mu$b for events that have both diffractive systems with 4.7$<$$|\eta|_{min}$$<$6.5. The values for the sub-categories are summarized in Table~\ref{results}. The measured cross-sections are between the Pythia and Phojet predictions for corresponding $\eta$ ranges.

We are grateful to the beam optics development team for the design and the successful commissioning
of the high $\beta^{*}$ optics and to the LHC machine coordinators for scheduling the dedicated fills. We thank P. Anielski, M. Idzik, I. Jurkowski, P. Kwiecien, R. Lazars, B. Niemczura for their help in software development. This work was supported by the institutions listed on the front page and partially also by NSF (US), the Magnus Ehrnrooth foundation (Finland), the Waldemar von Frenckell foundation (Finland), the Academy of Finland, the Finnish Academy of Science and Letters (The Vilho, Yrj\"o and Kalle V\"ais\"al\"a Fund), the OTKA grant NK 101438, 73143 (Hungary) and the NKTH-OTKA grant 74458 (Hungary).

\begin{widetext}

\begin{table}
\caption{Expected number of background events and observed number of data events passing the signal event selection 2T2+0T1. 
\label{bckg}}
\begin{ruledtabular}
\begin{tabular}{lccccc}
& ${\rm I_{track}}$ & ${\rm D11_{track}}$ & ${\rm D22_{track}}$ & ${\rm D12_{track}}$ & ${\rm D21_{track}}$ \\
\hline
ND & 829$\pm$239 & 672$\pm$100 & 28$\pm$22 & 115$\pm$16 & 109$\pm$23 \\
SD & 1,588$\pm$381 & 895$\pm$321 & 80$\pm$76 & 303$\pm$95 & 291$\pm$77 \\
CD & 7$\pm$3 & 5$\pm$3 & 1$\pm$1 & 1$\pm$1 & 1$\pm$1\\
\hline
Total expected background & 2,424$\pm$450 & 1,572$\pm$336 & 109$\pm$79 & 419$\pm$96 & 400$\pm$80 \\
\hline
Data & 8,214 & 5,261 & 375 & 1,350 & 1,386 \\
\end{tabular}
\end{ruledtabular}
\end{table}

\end{widetext}

\begin{widetext}

\begin{table}
\caption{Summary of statistical and systematic uncertainties ($\mu$b). \label{systematics}}
\begin{ruledtabular}
\begin{tabular}{lccccc}
& I & D11 & D22 & D12 & D21 \\
\hline
Statistical & 1.5 & 1.1 & 0.7 & 0.9 & 0.9 \\
Background estimate &  9.0 & 6.0 & 3.5 & 2.7 & 2.2 \\
Trigger efficiency & 2.1 & 1.2 & 1.0 & 0.9 & 0.9 \\
Pile-up correction & 2.4 & 2.1 & 0.4 & 1.1 & 1.0 \\
T1 multiplicity & 7.0 & 3.9 & 0.7 & 1.6 & 1.7 \\
Luminosity & 4.7 & 2.6 & 0.5 & 1.1 & 1.1 \\
Acceptance & 14.7 & 14.1 & 2.6 & 2.0 & 2.0 \\
True $\eta_{min}$ & 15.4 & 11.0 & 1.5 & 2.9 & 2.9 \\
\hline
Total uncertainty & 24.8 & 19.6 & 4.8 & 5.1 & 4.9 \\
\end{tabular}
\end{ruledtabular}
\end{table}

\begin{table}
\caption{Double diffractive cross-section measurements ($\mu$b) in the forward region. Both visible and true $\eta_{min}$ corrected cross-sections are given. The latter is compared to Pythia and Phojet predictions. Pythia estimate for total $\sigma_{DD}$=8.1~mb and Phojet estimate $\sigma_{DD}$=3.9~mb. \label{results}}
\begin{ruledtabular}
\begin{tabular}{lccccc}
& ${\rm I_{track}}$ & ${\rm D11_{track}}$ & ${\rm D22_{track}}$ & ${\rm D12_{track}}$ & ${\rm D21_{track}}$ \\
\hline
Visible & 131$\pm$22 & 58$\pm$14 & 20$\pm$8 & 31$\pm$5 & 34$\pm$5 \\
\hline
\hline
& I & D11 & D22 & D12 & D21 \\
\hline
True $\eta_{min}$ & 116$\pm$25 & 65$\pm$20 & 12$\pm$5 & 26$\pm$5 & 27$\pm$5  \\
\hline
Pythia true $\eta_{min}$ & 159 & 70 & 17 & 36 & 36 \\
Phojet true $\eta_{min}$ & 101 & 44 & 12 & 23 & 23 \\
\end{tabular}
\end{ruledtabular}
\end{table}

\end{widetext}

\bibliography{basename of .bib file}

\begin{thebibliography}{1}
\bibitem{pseudorapidity}$\eta$=-ln[tan($\theta$/2)] where $\theta$ is the polar angle.
\end{thebibliography}

\end{document}